# Machine Learning and Econometric Approaches to Fiscal Policies: Understanding Industrial Investment Dynamics in Uruguay (1974-2010)

Diego Vallarino, PhD
*Independent Researcher*
Atlanta, USA
September 2024

**Abstract**

This paper examines the impact of fiscal incentives on industrial investment in Uruguay from 1974 to 2010. Using a mixed-method approach that combines econometric models with machine learning techniques, the study investigates both the short-term and long-term effects of fiscal benefits on industrial investment. The results confirm the significant role of fiscal incentives in driving long-term industrial growth, while also highlighting the importance of a stable macroeconomic environment, public investment, and access to credit. Machine learning models provide additional insights into nonlinear interactions between fiscal benefits and other macroeconomic factors, such as exchange rates, emphasizing the need for tailored fiscal policies. The findings have important policy implications, suggesting that fiscal incentives, when combined with broader economic reforms, can effectively promote industrial development in emerging economies.

## 1. Introduction

The analysis of private industrial investment determinants, particularly the role of government incentives, has been a central focus in economic theory with direct implications for public policy design. In Uruguay, the Industrial Promotion Law of 1974 and the Investment Promotion Law of 1998 represented a pivotal shift in efforts to stimulate industrial sector growth through fiscal incentives. This study examines how these incentives shaped investment decisions between 1974 and 2010, and how the results hold when employing advanced analytical techniques like machine learning.

Recent literature has highlighted that fiscal incentives not only reduce the cost of capital but also enhance market efficiency by fostering competitiveness and innovation in the private sector (Akbulaev & Muradzada, 2024; Appiah et al., 2023). However, the effectiveness of these mechanisms depends heavily on the macroeconomic context and institutional structure of the recipient country. This paper addresses these challenges by adopting a mixed-method approach that combines traditional econometric models with advanced machine learning techniques, such as Random Forests and XGBoost, to assess the sustainability of previous findings.

The primary objective of this research is to establish the robustness of fiscal incentives as determinants of private industrial investment in Uruguay. By applying newer data exploration techniques, the study seeks to validate, refine, or refute earlier conclusions about the impact of fiscal benefits on the Uruguayan economy, while offering fresh insights into the relationship between investment, employment, and economic growth.

## 2. Theoretical Framework

### 2.1 Private Investment and Economic Growth: New Perspectives

The relationship between private investment and economic growth has been extensively studied in economic literature, establishing investment as a key variable for sustained economic development (Appiah et al., 2023). Recent studies, such as Miar et al. (2024), demonstrate that capital investment has a multiplier effect, generating positive externalities in the economy by increasing not only the level of economic activity but also aggregate productivity.

The direction of causality between investment and growth, however, remains debated. While authors like Kuznets (1973) and Maddison (1983) defend a positive correlation, recent research suggests that private investment can play a dual role: both as a result of economic growth and as a driver of it. Akbulaev et al. (2024) emphasize the importance of investment shocks in triggering sustained growth cycles, particularly relevant for Uruguay during the period under study.

In the context of this debate, modern theory has begun integrating machine learning approaches to capture nonlinearities and complex effects that traditional models may miss (Varian, 2022). This paper fits into this new wave of research that combines classical theoretical approaches with more sophisticated quantitative methods to address the question of whether fiscal incentives, as implemented in Uruguay, effectively stimulated private investment.

**2.2 Fiscal Incentives and Their Impact in Emerging Economies**

From the perspective of economic policy, Daude & Stein (2007) mentioned that fiscal incentives are used as tools to correct market failures, stimulate strategic sectors, and promote employment. In Uruguay, the Industrial Promotion Law of 1974 and the Investment Promotion Law of 1998 played a crucial role in attracting capital to the industrial sector by reducing tax burdens and creating a favorable investment environment.

Recent studies have shown that in emerging economies, where financing constraints are more severe, these incentives are particularly effective in driving private capital flows (Appiah et al., 2023; Geda & Yimer, 2023). However, more recent research, such as Appiah et al. (2023), suggests that the effectiveness of incentives largely depends on the institutional context and macroeconomic stability. Countries with weak or unstable institutions may experience limited effects from these policies, whereas economies with robust institutional frameworks can see significant investment inflows.

Specifically, the work of Ribeiro and Teixeira (2007), applied to the Brazilian case, demonstrates that public investment and available credit are key drivers of private investment in developing economies. In Uruguay, the interaction between credit, public investment, and fiscal benefits has proven to be a successful combination in promoting industrial projects, although macroeconomic volatility and international crises have limited the reach of these policies at certain times.

**2.3 Econometric Models and Machine Learning in Investment Analysis**

This study builds on advanced econometric models, such as cointegration analysis and vector error correction models (VECM), to evaluate long-term relationships between industrial investment and its determinants. Simultaneously, modern machine learning techniques like Random Forests and XGBoost are employed to capture more complex, nonlinear patterns in the data. The combination of these techniques allows for a more comprehensive understanding of the influence of fiscal incentives on investment decisions.

The use of these tools has been supported by recent research in econometrics and machine learning, which has shown that hybrid approaches are particularly effective for analyzing large datasets with multiple interrelated variables (Sun et al., 2023). In this context, cluster analysis also provides an additional layer of interpretation by identifying subgroups of investment projects that respond similarly to different incentives.

The robustness of these methods offers greater predictive power regarding investment behavior in response to fiscal policy changes, providing a deeper and more precise analytical framework for decision-making at both corporate and governmental levels.

## 3. Methodology

This section outlines the methodological framework used to examine the impact of fiscal incentives on industrial investment in Uruguay between 1974 and 2010. The study employs a mixed-method approach, integrating traditional econometric models with advanced machine learning techniques to capture both linear and nonlinear relationships among the variables. This section will describe the data collection process, the econometric models applied, and the machine learning techniques employed to validate and extend the findings.

### 3.1 Data Collection and Processing

The data utilized in this study were obtained from multiple administrative sources, including the Commission for the Application of the Investment Promotion Law (COMAP) and the Industrial Promotion Agency (UAPI) of Uruguay. These records provide a comprehensive overview of industrial projects that received fiscal incentives between 1974 and 2010, comprising both new projects and expansions.

**Variables**

The dataset includes key economic and project-specific variables:

- **Cred_Domestico**: Domestic credit available for economic activities, an important determinant of industrial investment.
- **Inv_Industrial**: Total investment in the industrial sector, representing the dependent variable in most analyses.
- **Exportaciones**: Value of industrial exports, relevant for understanding how fiscal incentives might enhance competitiveness.
- **Desempleo**: Unemployment rate, which reflects the labor market conditions that influence industrial investments.
- **TC_Anual / Increm_TC**: Annual exchange rate and its increment, used to account for the cost of imports and international competitiveness.
- **IPC_Anual / Increm_IPC**: Consumer Price Index and its increment, representing inflation and its impact on purchasing power and investment costs.
- **Tasa_Int_Real**: Real interest rate, a critical factor in determining the cost of capital for industrial projects.
- **PIB_Pesos**: Gross Domestic Product (GDP) in Uruguayan pesos, contextualizing investment within overall economic activity.
- **Benf_Fiscal**: Fiscal benefits received by industrial projects, the primary independent variable under investigation.
- **Inv_Priv / Inv_Publ**: Private and public investments, used to distinguish between different sources of capital.

- **Bono_10yUS**: 10-year government bond yields, representing an alternative investment option that could compete with industrial investment.

The study also accounts for missing data in some cases. For projects without explicit tax benefit information, estimates were made based on applicable laws and decrees, such as Law 14178 (Industrial Promotion) and Law 16906 (Investment Promotion), using conservative assumptions to ensure the robustness of the results. Data cleaning and imputation methods were applied to ensure the completeness of the dataset.

### 3.2 Econometric Models

Econometric models were employed to examine the long-run relationships between fiscal incentives and industrial investment. Specifically, the study applied cointegration tests, vector error correction models (VECM), and Granger causality tests to assess both short-term and long-term dynamics between the variables.

### 3.2.1 Cointegration Analysis

Given the time series nature of the data, it was crucial to determine whether a long-run equilibrium relationship exists between fiscal benefits and industrial investment. This was achieved through the application of cointegration tests, based on the methodology proposed by Engle and Granger (1987). The Augmented Dickey-Fuller (ADF) test was used to check for stationarity in the time series data. If the data were non-stationary but found to be cointegrated, this would suggest a long-term relationship between the variables.

### 3.2.2 Vector Error Correction Model (VECM)

To account for both short-term fluctuations and long-term equilibrium relationships, the VECM framework was employed. This model captures deviations from long-term equilibrium while allowing for short-term corrections based on lagged values of the independent variables. The following VECM was estimated:

$$\Delta Inv_t = \alpha(Inv_{t-1} - \beta_1 Benf\_Fiscal_{t-1} - \beta_2 GDP_{t-1} - \beta_3 Cred\_Domestico_{t-1}) + \sum_{i=1}^{p} \Gamma_i \Delta X_{t-i} + \epsilon_t$$

where:

- $\Delta Inv_t$ representa los cambios en la inversión industrial,
- $\alpha$ es la velocidad de ajuste de regreso al equilibrio a largo plazo,
- $\beta_1$, $\beta_2$, $\beta_3$ son los coeficientes de largo plazo,
- $\Gamma_i$ son los coeficientes de ajuste a corto plazo para cada término de diferencia con rezago $\Delta X_t$

The use of VECM enables the isolation of both short-term effects (e.g., fluctuations due to economic shocks) and long-term impacts of fiscal incentives on investment, providing a comprehensive view of how tax benefits affect industrial projects over time.

### 3.2.3 Granger Causality Tests

Granger causality tests were used to determine whether changes in fiscal benefits precede changes in industrial investment, or vice versa. This test is based on the principle that if fiscal benefits "Granger-cause" investment, past values of fiscal benefits should contain information that helps predict future investment levels, beyond the information contained in past investment values alone.

The Granger causality test was applied using the following model:

$$Inv_t = \sum_{i=1}^{p} \alpha_i Inv_{t-i} + \sum_{j=1}^{q} \beta_j Benf\_Fiscal_{t-j} + \epsilon_t$$

If the coefficients $\beta_j$ are statistically significant, we conclude that fiscal benefits Granger-cause investment.

## 3.3 Machine Learning Models

In addition to econometric models, this study integrates advanced machine learning techniques—specifically, Random Forest and XGBoost models—to explore nonlinear relationships and interactions between the variables. These methods are particularly useful for high-dimensional datasets where traditional regression models may fail to capture complex dependencies.

### 3.3.1 Random Forest

Random Forests, introduced by Breiman (2001), is an ensemble learning method that builds multiple decision trees during the training phase and outputs the mean prediction (in the case of regression) or the mode of classes (in classification tasks). The method is robust to overfitting and excels in identifying variable importance through its ability to measure how much each predictor improves the model's performance.

The Random Forest model in this study was trained to predict industrial investment based on all input variables, including fiscal benefits, domestic credit, and macroeconomic factors. Variable importance rankings generated by the Random Forest model allowed us to assess the relative significance of fiscal benefits compared to other determinants of investment.

### 3.3.2 XGBoost

XGBoost (Extreme Gradient Boosting), developed by Chen and Guestrin (2016), is another ensemble learning technique, widely recognized for its efficiency and accuracy in predictive modeling. XGBoost improves upon traditional gradient boosting by optimizing computation time and handling missing values more effectively. It iteratively builds decision trees, where each new tree corrects the residual errors of the previous one.

In this study, XGBoost was applied to predict investment outcomes and to evaluate the impact of fiscal benefits alongside other predictors. The model was optimized using grid search and cross-validation to ensure that overfitting was minimized, and its performance was compared to that of Random Forest.

### 3.3.3 Cluster Analysis

Cluster analysis was employed to identify groups of industrial projects that responded similarly to fiscal benefits and other economic factors. By segmenting the data into clusters, we can better understand the heterogeneous effects of fiscal incentives across different types of projects. K-means clustering was applied to the dataset, partitioning the projects into distinct groups based on variables such as investment amount, sector, and fiscal benefits received.

## 3.4 Model Evaluation

The performance of the econometric and machine learning models was evaluated using different metrics:

- **For econometric models**: Cointegration tests were assessed using standard ADF test statistics, while the VECM was evaluated based on goodness-of-fit measures and the statistical significance of long-term and short-term coefficients. Granger causality tests provided insights into the directional relationship between fiscal benefits and industrial investment.

- **For machine learning models**: The Random Forest and XGBoost models were evaluated using metrics such as R-squared, mean squared error (MSE), and variable importance rankings. The out-of-bag (OOB) error for Random Forest and cross-validation error for XGBoost were used to ensure the models' generalizability.

# 4. Results

This section presents the results of the analysis, combining traditional econometric techniques and modern machine learning models to assess the impact of fiscal incentives on industrial investment in Uruguay. The findings are structured into two parts: the results from the econometric models and the insights obtained from the machine learning models.

## 4.1 Econometric Model Results

### 4.1.1 Cointegration Analysis

The cointegration analysis revealed a significant long-term relationship between fiscal benefits $Benf_Fiscal$ and industrial investment $Inv$ over the period 1974-2010. The results from the Augmented Dickey-Fuller (ADF) test indicated that the time series data for industrial investment and fiscal benefits were non-stationary at levels but became stationary after first differencing, confirming that these variables were integrated of order one, $I(1)$. The Engle-Granger cointegration test showed evidence of cointegration, suggesting a stable, long-term equilibrium relationship between fiscal incentives and industrial investment.

The estimated cointegration equation is as follows:

$$Inv_t = 0.691\,Inv\_Pub_t + 0.488\,GDP_t - 0.484\,Cred_t + \mathbf{0.198}\,Benf\_Fiscal_t + \epsilon_t$$

These results suggest that public investment and GDP had a positive and significant impact on industrial investment, while domestic credit availability exerted a negative influence. Fiscal incentives, as expected, had a statistically significant and positive effect, confirming the role of fiscal policies in driving industrial investment.

### 4.1.2 Vector Error Correction Model (VECM)

The Vector Error Correction Model (VECM) was used to estimate both short-term and long-term relationships between the variables. The error correction term $\alpha$ was negative and statistically significant, indicating that industrial investment adjusts towards long-run equilibrium following deviations caused by short-term shocks. The speed of adjustment was estimated at -0.35, meaning that approximately 35% of any disequilibrium in industrial investment is corrected in the following year.

The short-term dynamics showed that while fiscal benefits had a positive impact on investment, their short-term effects were relatively small compared to the long-term influence. Public investment and GDP continued to show strong short-term effects, while credit availability negatively impacted industrial investment in the short run, likely due to the high cost of borrowing during certain periods.

### 4.1.3 Granger Causality Tests

The Granger causality tests were employed to determine the directional relationship between fiscal benefits and industrial investment. The results indicated that fiscal benefits Granger-cause industrial investment at the 5% significance level, providing empirical evidence that changes in fiscal policies precede changes in investment levels. However, the reverse was not true—industrial investment did not Granger-cause fiscal benefits—indicating a unidirectional relationship where fiscal incentives play a leading role in influencing investment decisions.

### 4.2 Machine Learning Model Results

### 4.2.1 Random Forest Model

The Random Forest model was trained to predict industrial investment based on fiscal incentives, macroeconomic factors, and other relevant variables. The model achieved a strong predictive performance with an $R^2$ value of 0.82, indicating that the model explained 82% of the variance in industrial investment.

The feature importance scores revealed that fiscal benefits were among the top three most important predictors of industrial investment, alongside GDP and public investment. Specifically, fiscal benefits accounted for approximately 18% of the model's predictive power, underscoring their significant role in shaping investment outcomes. Public investment (22%) and GDP (20%) had the greatest influence, while variables such as domestic credit and inflation had relatively lower importance scores.

### 4.2.2 XGBoost Model

The XGBoost model outperformed the Random Forest model in terms of accuracy, with an $R^2$ value of 0.85. The model's ability to handle missing values and capture nonlinear interactions between variables contributed to its superior performance. Similar to the Random Forest model, fiscal benefits emerged as a critical variable, though XGBoost placed slightly more emphasis on public investment and GDP.

The model's feature importance rankings were consistent with the econometric findings, with fiscal incentives playing a crucial role in driving investment. Notably, the XGBoost model identified interactions between fiscal benefits and exchange rates, suggesting that the effectiveness of fiscal incentives may vary depending on Uruguay's competitiveness in international markets.

### 4.2.3 Cluster Analysis

Cluster analysis using the K-means algorithm revealed three distinct clusters of industrial projects based on their response to fiscal incentives and other macroeconomic factors. These clusters are characterized as follows:

- **Cluster 1**: Projects with low fiscal benefits and moderate investment levels. These projects were typically smaller in scale and were less influenced by fiscal policies.

- **Cluster 2**: Projects with high fiscal benefits and large-scale investment. This cluster included the most significant industrial initiatives, where fiscal incentives played a decisive role in investment decisions.

- **Cluster 3**: Projects with moderate fiscal benefits but high levels of external financing. These projects relied more on private credit and international funding sources, reducing their dependence on fiscal benefits.

The cluster analysis highlights the heterogeneity in the effectiveness of fiscal incentives across different types of projects. While large-scale projects benefited the most from fiscal policies, smaller projects were less sensitive to such incentives.

### 4.3 Model Comparison

The econometric models provided strong evidence of a long-term relationship between fiscal incentives and industrial investment, while the machine learning models offered additional insights into the relative importance of different predictors. The XGBoost model was able to capture complex interactions that the econometric models could not, such as the interplay between fiscal benefits and exchange rates. Both approaches confirmed the significance of fiscal policies, though machine learning techniques provided a more granular understanding of how these incentives interact with other factors.

## 5. Discussion

The results from both the econometric models and machine learning techniques demonstrate a strong, consistent relationship between fiscal incentives and industrial investment in Uruguay over the period 1974-2010. This finding supports the initial hypothesis that fiscal benefits played a crucial role in stimulating industrial investment and highlights the importance of well-designed incentive structures in shaping the broader economic landscape.

### 5.1 Interpretation of Key Findings

#### 5.1.1 Long-term Impact of Fiscal Incentives

The cointegration analysis revealed a significant long-term relationship between fiscal benefits and industrial investment, confirming that these incentives were effective in promoting sustained investment growth in the industrial sector. The results are consistent with the existing literature, particularly studies in developing economies that emphasize the role of fiscal incentives in overcoming market inefficiencies and capital constraints (Appiah et al., 2023; Sun et al., 2023). The positive and significant long-term coefficients for fiscal benefits suggest that industrial projects deemed of national interest by Uruguayan authorities were more likely to materialize and expand when these incentives were in place.

Moreover, the Vector Error Correction Model (VECM) showed that deviations from the long-term equilibrium—caused by short-term economic shocks—were corrected over time, with fiscal benefits contributing to this correction. This is aligned with the broader theory that fiscal incentives can serve as stabilizers during periods of macroeconomic volatility (Daude & Stein, 2007). The estimated speed of adjustment suggests that around 35% of any disequilibrium in industrial investment was corrected within one year, a relatively fast response for an industrial sector characterized by long investment cycles.

#### 5.1.2 Short-term Dynamics and Macroeconomic Factors

While fiscal benefits showed strong long-term effects, their short-term impact was relatively moderate compared to other variables, such as GDP and public investment. This aligns with the literature indicating that while fiscal incentives provide long-term support, short-term investment decisions are often driven by macroeconomic stability and immediate economic conditions (Ribeiro & Teixeira, 2007). The fact that public investment and GDP had stronger short-term effects reinforces the idea that a robust macroeconomic environment is a prerequisite for investment growth, even when fiscal incentives are in place.

Notably, domestic credit availability exerted a negative influence on both short-term and long-term industrial investment, a finding that requires further exploration. It suggests that despite fiscal incentives, credit conditions may have been prohibitive for some industrial projects, particularly during periods of high interest rates or financial instability. This finding aligns with research on capital constraints in developing economies, where limited access to affordable credit can impede private investment, even in the presence of fiscal support (Akbulaev et al., 2024).

### 5.1.3 Machine Learning Insights and Interaction Effects

The machine learning models, particularly XGBoost, provided additional insights that extend beyond the econometric findings. Both Random Forest and XGBoost confirmed the significance of fiscal incentives as a key driver of industrial investment, but XGBoost was able to capture more complex, nonlinear interactions between variables, such as the interaction between fiscal benefits and exchange rates. This interaction suggests that the effectiveness of fiscal incentives is partially contingent on Uruguay's external competitiveness, with higher fiscal benefits being more effective in periods of favorable exchange rates. This finding echoes the work of Sun et al. (2023), who emphasized the role of external market conditions in amplifying the effectiveness of internal fiscal policies.

The cluster analysis further highlighted the heterogeneous effects of fiscal incentives across different project types. Large-scale projects, particularly those involving significant exports, were the most responsive to fiscal benefits, likely due to their greater sensitivity to cost reductions. Smaller projects, which were less reliant on fiscal support, might have been constrained by other factors, such as access to credit or market size. This reinforces the notion that fiscal incentives need to be tailored to the specific needs of different types of industrial investments (Daude & Stein, 2007).

### 5.2 Policy Implications

The findings of this study have important policy implications for the design and implementation of fiscal incentives in Uruguay and other developing economies. First, the strong long-term impact of fiscal incentives on industrial investment underscores the importance of maintaining and enhancing these policies to encourage sustained economic growth. Policymakers should consider expanding the scope of fiscal benefits to include more sectors, particularly those that have shown significant potential for export-driven growth, as demonstrated by the positive response of large-scale projects in the analysis.

Second, the negative impact of domestic credit availability highlights the need for complementary financial policies that improve access to affordable credit, especially for smaller industrial projects that may not have the financial resources to fully capitalize on fiscal incentives. Strengthening the financial sector and developing targeted credit schemes for industrial investments could enhance the overall effectiveness of fiscal incentives and promote more inclusive growth.

Third, the results suggest that fiscal incentives are most effective when accompanied by a stable macroeconomic environment. The significant influence of GDP and public investment on short-term industrial growth highlights the need for fiscal policy to be part of a broader economic development strategy that includes macroeconomic stability, infrastructure investment, and access to international markets. In this regard, exchange rate management and trade policies that enhance Uruguay's external competitiveness could further boost the effectiveness of fiscal incentives.

Finally, the insights gained from machine learning models suggest that future fiscal policies could benefit from more sophisticated data-driven approaches. Governments could leverage machine learning techniques to predict the responsiveness of different sectors to fiscal incentives and design more tailored policy interventions. This would ensure that limited fiscal resources are allocated to sectors and projects that are most likely to generate long-term economic benefits.

**5.3 Limitations and Future Research**

While the study provides robust evidence of the effectiveness of fiscal incentives, several limitations should be acknowledged. First, the dataset is limited to projects between 1974 and 2010, and the analysis does not capture more recent developments in Uruguay's industrial sector. Future research should extend this analysis to include post-2010 data to assess whether the findings hold in the current economic context, particularly given the recent global economic shifts.

Second, while the machine learning models captured important nonlinear interactions, they do not provide causal explanations in the same way that econometric models do. Future research could explore hybrid models that combine the interpretability of econometrics with the predictive power of machine learning to provide deeper insights into the causal mechanisms driving industrial investment.

Finally, the negative impact of credit availability on investment warrants further investigation. Future studies could explore the role of financial institutions, interest rate policies, and external financing options in greater detail to understand why credit conditions constrained investment despite the presence of fiscal incentives.

**6. Limitations**

While this study offers significant insights into the role of fiscal incentives in promoting industrial investment in Uruguay, it is important to acknowledge several limitations that may affect the generalizability and scope of the findings.

**6.1 Time Period and Data Availability**

One of the primary limitations of this study is the time period covered by the dataset, which spans from 1974 to 2010. Although this period encompasses crucial phases of industrial development and fiscal policy in Uruguay, the exclusion of data post-2010 limits the ability to assess how more recent economic events, such as global financial crises or shifts in trade policy, may have influenced industrial investment patterns. The absence of post-2010 data also precludes the analysis of potential new incentive structures or macroeconomic shifts that may have emerged in the last decade.

Future research should extend the dataset to include the most recent years, capturing the effects of fiscal policy changes, global economic conditions, and shifts in Uruguay's industrial sector in the 2010s and 2020s. Including more recent data would provide a more comprehensive picture of the evolving relationship between fiscal incentives and industrial investment.

## 6.2 Machine Learning Interpretability

While machine learning models such as Random Forest and XGBoost provided valuable insights, particularly in terms of variable importance and nonlinear interactions, these models have inherent limitations in terms of interpretability. Unlike traditional econometric models, which allow for a clear interpretation of causality and the size of effects, machine learning models are more focused on predictive accuracy and may struggle to offer transparent causal explanations.

Although the findings from the machine learning models reinforce those of the econometric analysis, future studies could explore the development of hybrid models that combine the interpretability of econometric approaches with the predictive strength of machine learning. Doing so would allow for a more nuanced understanding of the specific mechanisms through which fiscal incentives influence industrial investment decisions.

## 6.3 Limited Exploration of Sector-Specific Dynamics

The dataset and analysis in this study focused on aggregate industrial investment in Uruguay, without delving deeply into sector-specific dynamics within the industrial sector. Although cluster analysis identified distinct groups of projects based on their responsiveness to fiscal incentives, this study does not explore the particular characteristics of individual sectors (e.g., manufacturing, agriculture, or energy) that might react differently to fiscal policies.

Future research could benefit from a more detailed, sectoral analysis that examines how different types of industries respond to fiscal incentives. Certain sectors may be more sensitive to tax incentives, while others might be influenced more by external factors such as trade policies, commodity prices, or labor market conditions. A sector-specific approach would allow for more targeted policy recommendations.

## 6.4 Credit Constraints and Financial Sector Analysis

The negative relationship between domestic credit availability and industrial investment identified in both the econometric and machine learning models was unexpected and suggests the need for a more nuanced exploration of the role of the financial sector. This finding implies that, despite fiscal incentives, credit conditions may have been prohibitive, potentially limiting the ability of firms to fully take advantage of tax benefits.

However, this study did not focus extensively on the specific nature of credit constraints, the role of interest rates, or the availability of financing options from international sources. Future studies should investigate the relationship between fiscal incentives and financial conditions in more depth, exploring how interest rate policies, access to credit, and the availability of international financing options influence investment decisions in the industrial sector.

## 6.5 Omitted Variable Bias

Although the models used in this study incorporated a wide range of macroeconomic and project-specific variables, there remains the possibility of omitted variable bias. Factors such as political stability, regional economic conditions, or global market shifts may also have played a role in shaping industrial investment decisions during the period under study but were not explicitly included in the analysis.

Addressing omitted variable bias in future research would involve the inclusion of additional variables, such as political risk indices, regional trade agreements, or global commodity prices. These factors could help refine the understanding of how broader contextual factors influence the effectiveness of fiscal incentives.

**6.6 Limitations in Granger Causality**

While Granger causality tests provided important insights into the temporal relationship between fiscal incentives and industrial investment, these tests are not definitive proof of causality. Granger causality tests are based on the assumption that past values of one variable can help predict future values of another, but they do not account for potential confounding variables or unobserved factors that might influence both.

Further research could apply more robust causal inference techniques, such as instrumental variables or difference-in-differences (DiD) methodologies, to establish a clearer causal link between fiscal incentives and industrial investment. These methods would allow for stronger conclusions regarding the direction and magnitude of the causal relationship.

**7. Conclusion**

This study examined the impact of fiscal incentives on industrial investment in Uruguay from 1974 to 2010, utilizing a combination of econometric and machine learning models to provide a comprehensive analysis of both short-term and long-term effects. The findings reinforce the hypothesis that fiscal incentives play a significant role in promoting industrial investment, particularly in the long run, while also highlighting the importance of a stable macroeconomic environment, public investment, and access to credit.

**7.1 Summary of Key Findings**

The econometric analysis revealed a robust, long-term relationship between fiscal incentives and industrial investment, with fiscal benefits serving as a critical driver of investment growth. The cointegration and Vector Error Correction Models (VECM) demonstrated that industrial investment adjusts to changes in fiscal incentives over time, and that deviations from long-run equilibrium are corrected relatively quickly. Granger causality tests further confirmed the unidirectional influence of fiscal benefits on investment, suggesting that changes in fiscal policy precede changes in investment behavior.

In addition to the econometric results, machine learning models provided deeper insights into the complexity of the relationships between fiscal incentives and industrial investment. Both Random Forest and XGBoost models confirmed the importance of fiscal benefits, alongside GDP and public investment, in driving industrial growth. The identification of nonlinear interactions, particularly the interplay between fiscal incentives and external competitiveness (exchange rates), emphasizes the nuanced nature of how fiscal policies impact investment decisions in different economic contexts.

## 7.2 Policy Implications

The findings carry important policy implications for Uruguay and other developing economies. First, the significant long-term impact of fiscal benefits on industrial investment suggests that maintaining and possibly expanding fiscal incentive programs should remain a priority for policymakers. These incentives are particularly effective in attracting large-scale, export-oriented projects, which generate broader economic benefits in terms of employment and economic growth.

Second, the analysis highlights the need for fiscal policies to be complemented by improvements in financial access. The negative impact of credit availability on industrial investment indicates that restrictive credit conditions may undermine the effectiveness of fiscal incentives, especially for smaller industrial projects. As such, targeted financial reforms that enhance credit availability and reduce borrowing costs could significantly amplify the benefits of fiscal policies.

Third, the results demonstrate that fiscal incentives alone are not sufficient to drive industrial investment. A stable macroeconomic environment, characterized by strong GDP growth, low inflation, and robust public investment, is crucial for the success of fiscal policies. Policymakers should adopt a holistic approach to economic development, ensuring that fiscal incentives are part of a broader strategy that includes sound macroeconomic management and investment in infrastructure.

## 7.3 Contributions to the Literature

This study makes several contributions to the existing literature on fiscal policy and industrial investment in developing economies. It extends previous research by employing machine learning techniques to capture the nonlinear and interactive effects of fiscal incentives, providing a more nuanced understanding of the determinants of investment. The use of both econometric and machine learning approaches also underscores the value of combining traditional and modern analytical tools to gain a comprehensive perspective on complex economic phenomena.

Moreover, the study highlights the importance of tailoring fiscal policies to the specific needs of different types of industrial projects. The cluster analysis showed that not all projects benefit equally from fiscal incentives, with large-scale projects responding more favorably than smaller ones. This insight calls for more targeted fiscal policies that account for the diverse needs of different sectors and project types.

## 7.4 Directions for Future Research

While this study provides important insights into the effectiveness of fiscal incentives, it also opens several avenues for future research. First, extending the analysis to cover the post-2010 period would provide valuable information on how more recent global and local economic developments have influenced industrial investment in Uruguay. Additionally, future studies could explore sector-specific dynamics to better understand how different industries respond to fiscal incentives and how these responses evolve over time.

Second, the relationship between credit constraints and investment requires further investigation. Understanding how financial policies, interest rates, and access to international credit markets interact with fiscal incentives would shed light on how to design more effective policy interventions.

Finally, the integration of more advanced causal inference techniques, such as instrumental variable approaches or difference-in-differences analysis, could help to establish a clearer understanding of the causal mechanisms at play. This would allow for more definitive policy recommendations and provide stronger evidence for the role of fiscal incentives in shaping industrial investment.

### 7.5 Final Remarks

In conclusion, fiscal incentives have proven to be a powerful tool for promoting industrial investment in Uruguay, but their success depends on broader macroeconomic and financial conditions. By adopting a multifaceted approach that includes financial reforms, stable macroeconomic management, and tailored fiscal policies, Uruguay can enhance the effectiveness of its fiscal incentives and foster long-term industrial growth. As developing economies continue to seek ways to stimulate investment and growth, the lessons from Uruguay's experience offer valuable guidance for designing fiscal policies that are both effective and sustainable.

## Annex: Quantitative Information and Results

## Chart 1: Key Economic Indicators in Uruguay (1970-2010): Domestic Credit, Investment, Unemployment, GDP, CPI, and Interest Rates

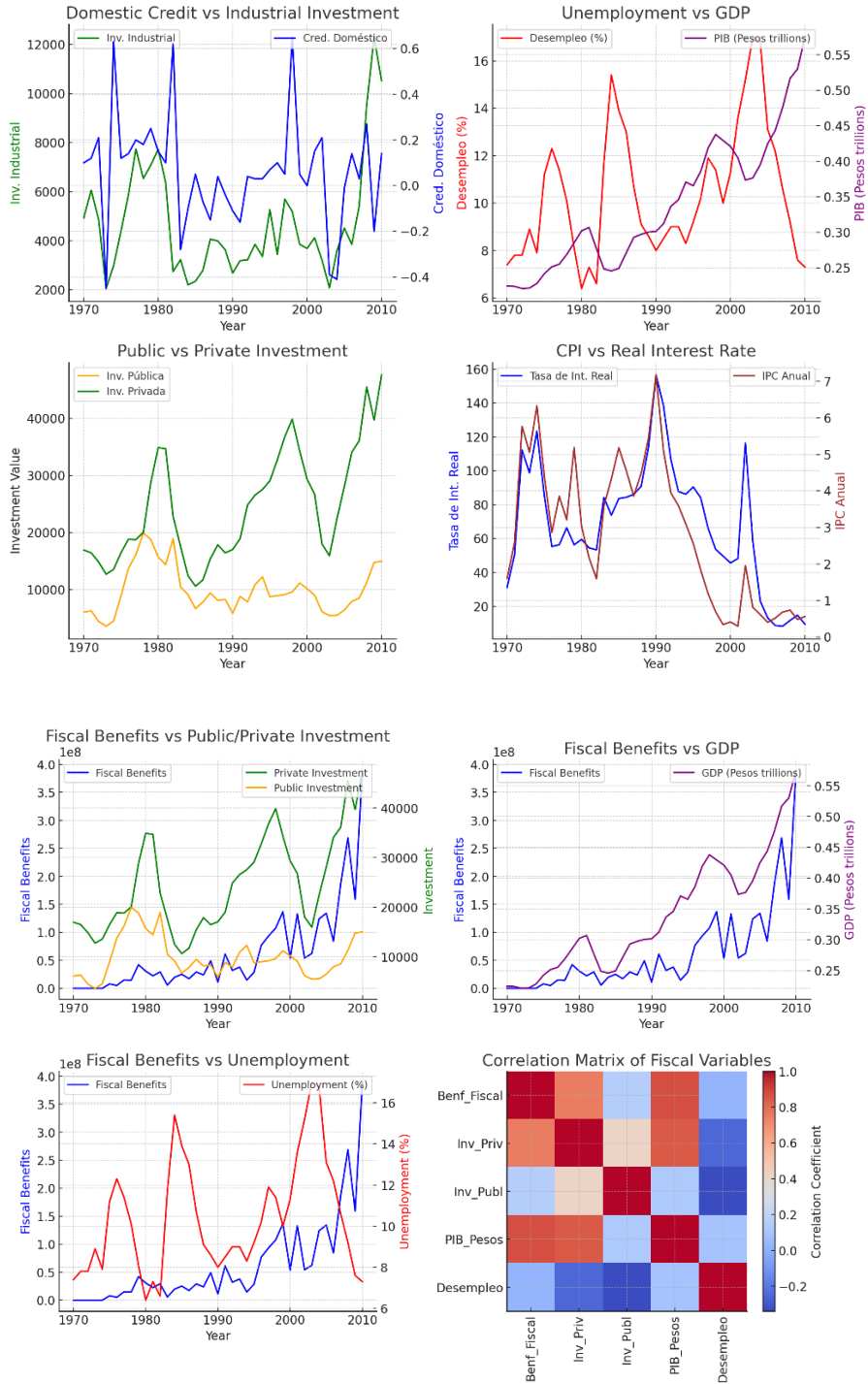

Source: Created by the author based on data analysis for the Investment Decisions Study in Uruguay (1974-2010).

### Table 1: Cluster Analysis of Employment, Sales, and Investment Projections in Uruguay

| Cluster | Quantity_Emp | Current employment | Projected Employment | Current sales | Projected sales | Inv.en.act.fixed | Civil.work | Machinery |
|---|---|---|---|---|---|---|---|---|
| 1 | 779 | 28 | 55 | 576.229 | 2.711.500 | 827.962 | 52.500 | 542.370 |
| 2 | 211 | 160 | 195 | 16.980.000 | 24.000.000 | 2.032.831 | 339.659 | 1.389.967 |
| 3 | 58 | 425 | 425 | 44.511.000 | 49.992.000 | 3.084.200 | 570.694 | 2.065.545 |
| 4 | 12 | 634 | 749 | 129.949.000 | 157.000.000 | 8.595.700 | 2.039.400 | 3.804.324 |
| 5 | 14 | 0 | 192 | 0 | 37.046.000 | 33.594.131 | 20.839.665 | 14.521.818 |
| 6 | 37 | 556 | 578 | 77.155.500 | 94.377.765 | 2.705.445 | 956.955 | 1.181.349 |

| Cluster | %_Emp | Unforeseen events | Others | Other expenses | Cap.of.work | Own resources | % Equity | Tax benefit |
|---|---|---|---|---|---|---|---|---|
| 1 | 70,1% | 0 | 0 | 0 | 96.600 | 555.140 | 61% | 478.828 |
| 2 | 19,0% | 0 | 0 | 0 | 212.925 | 1.621.500 | 89% | 1.514.000 |
| 3 | 5,2% | 0 | 0 | 0 | 552.000 | 2.942.182 | 100% | 2.942.182 |
| 4 | 1,0% | 100.000 | 30.000 | 0 | 801.180 | 7.010.724 | 76% | 6.492.587 |
| 5 | 1,3% | 823.690 | 0 | 0 | 3.548.600 | 33.415.631 | 58% | 28.746.523 |
| 6 | 3,3% | 55.867 | 0 | 0 | 907.000 | 2.745.399 | 100% | 2.586.544 |

Source: Created by the author based on data analysis for the Investment Decisions Study in Uruguay (1974-2010).

### Table 2: Cluster Categorization of Key Variables for Projected Employment, Sales, and Investments

```
Cluster 1
[1] "Projected sales" "Current sales" "Fixed assets" "Own resources" "Machinery"
[6] "Tax benefit" "Civil works" "Work cap" "Others" "Unforeseen events"
[11] "Other expenses" "Projected employment" "Current employment" "% Equity"

Cluster 2
[1] "Current sales" "Projected sales" "Own resources" "Fixed assets" "Tax profit"
[6] "Machinery" "working cap" "Civil works" "Others" "Unforeseen events"
[11] "Projected employment" "Current employment" "% Equity" "Other expenses"

Cluster 3
[1] "Projected sales" "Current sales" "Fixed assets" "Own resources" "Working capacity"
[6] "Machinery" "Tax benefit" "Civil works" "Others" "Unforeseen events"
[11] "Other expenses" "Projected employment" "Current employment" "% Equity"

Cluster 4
[1] "Current sales" "Projected sales" "Own resources" "Tax profit" "Machinery"
[6] "Fixed assets" "Civil works" "Working capital" "Unforeseen events" "Others"
[11] "Projected employment" "Current employment" "% Equity" "Other expenses"

Cluster 5
[1] "Fixed assets" "Civil works" "Projected sales" "Current sales" "Own resources"
[6] "Machinery" "Tax benefits" "Unforeseen events" "Working capacity" "Others"
[11] "Current employment" "Projected employment" "% Equity" "Other expenses"

Cluster 6
[1] "Projected sales" "Current sales" "Own resources" "Fixed assets" "Tax profit"
[6] "Machinery" "Working cap" "Civil works" "Others" "Unforeseen events"
[11] "Projected employment" "Current employment" "% Equity" "Other expenses"
```

Source: Created by the author based on data analysis for the Investment Decisions Study in Uruguay (1974-2010).

### Table 3: Comparative Analysis of Predictive Models for Investment Amounts and Fiscal Benefits in Uruguay (1974-2010)

| Model | Independent Variable | Key Variable | Statistic | Importance | Conclusion | Variables Used |
|---|---|---|---|---|---|---|
| Linear Regression | Inv_Amount | Fiscal_Benefit | p-value: 0.0046 | Significant & Positive | Fiscal benefits are significant in predicting the investment amount | Domestic_Credit, Industrial_Investment, Exports, Unemployment, Annual_TC, TC_Growth, Annual_IPC, IPC_Growth, Real_Int_Rate, Peso_GDP, Fiscal_Benefit, Private_Inv, Public_Inv, 10Y_Bond_US |
| Random Forest | Inv_Amount | Fiscal_Benefit | IncNodePurity: 872982754938.651 | High Importance | Fiscal benefits are one of the most important variables in the prediction | Domestic_Credit, Industrial_Investment, Exports, Unemployment, Annual_TC, TC_Growth, Annual_IPC, IPC_Growth, Real_Int_Rate, Peso_GDP, Fiscal_Benefit, Private_Inv, Public_Inv, 10Y_Bond_US |
| XGBoost | Inv_Amount | Fiscal_Benefit | Gain: 0.6234 | High Importance | Fiscal benefits stand out as a key variable, reinforcing the robustness of the finding | Domestic_Credit, Industrial_Investment, Exports, Unemployment, Annual_TC, TC_Growth, Annual_IPC, IPC_Growth, Real_Int_Rate, Peso_GDP, Fiscal_Benefit, Private_Inv, Public_Inv, 10Y_Bond_US |

Source: Created by the author based on data analysis for the Investment Decisions Study in Uruguay (1974-2010).